\documentclass[10pt,a4paper,english,conference]{IEEEtran}
\usepackage[T1]{fontenc}
\usepackage{verbatim}
\usepackage{float}
\usepackage{textcomp}
\usepackage{mathrsfs}
\usepackage{amsmath}
\usepackage{amsthm}
\usepackage{amssymb}
\usepackage{graphicx}
\usepackage{setspace}

\makeatletter

\pdfpageheight\paperheight
\pdfpagewidth\paperwidth

\floatstyle{ruled}
\newfloat{algorithm}{tbp}{loa}
\providecommand{\algorithmname}{Algorithm}
\floatname{algorithm}{\protect\algorithmname}

\theoremstyle{plain}
\newtheorem{thm}{\protect\theoremname}
\theoremstyle{plain}
\newtheorem{lem}[thm]{\protect\lemmaname}

\usepackage{subfigure}
\usepackage{epstopdf}
\usepackage{cite}
\usepackage{citesort}
\usepackage{balance}

\makeatother

\usepackage{babel}
\providecommand{\lemmaname}{Lemma}
\providecommand{\theoremname}{Theorem}

\begin{document}

\title{Ultra-Dense Networks: A New Look at \\
the Proportional Fair Scheduler
}
\begin{singlespace}

\author{\noindent {\normalsize{}Ming Ding$^{\ddagger}$, David L$\acute{\textrm{o}}$pez
P$\acute{\textrm{e}}$rez$^{\dagger}$}\emph{\normalsize{}, }{\normalsize{}Amir
H. Jafari$^{\ast}$, Guoqiang Mao$^{\nparallel}{}^{\ddagger}$}\emph{\normalsize{},
}{\normalsize{}Zihuai Lin}\textit{\normalsize{}$^{\mathparagraph}$}\emph{\normalsize{}}\\
\textit{\footnotesize{}$^{\ddagger}$Data61, Australia, $^{\dagger}$Nokia
Bell Labs, Ireland}\\
\textit{\footnotesize{}$^{\nparallel}$School of Computing and Communication,
University of Technology Sydney, Australia}\\
\textit{\footnotesize{}$^{\ast}$Dept. of Electronic \& Electrical
Engineering, University of Sheffield, UK}\\
\textit{\footnotesize{}$^{\mathparagraph}$The University of Sydney,
Australia}
}
\end{singlespace}
\maketitle
\begin{abstract}
In this paper, we theoretically study the proportional fair (PF) scheduler
in the context of ultra-dense networks (UDNs). Analytical results
are obtained for the coverage probability and the area spectral efficiency
(ASE) performance of dense small cell networks (SCNs) with the PF
scheduler employed at base stations (BSs). The key point of our analysis
is that the typical user is no longer a random user as assumed in
most studies in the literature. Instead, a user with the maximum PF
metric is chosen by its serving BS as the typical user. By comparing
the previous results of the round-robin (RR) scheduler with our new
results of the PF scheduler, we quantify the loss of the multi-user
diversity of the PF scheduler with the network densification, which
casts a new look at the role of the PF scheduler in UDNs. Our conclusion
is that the RR scheduler should be used in UDNs to simplify the radio
resource management (RRM)%
\footnote{To appear in IEEE GLOBECOM2017. 1536-1276 © 2015 IEEE. Personal use is permitted, but republication/redistribution requires IEEE permission. Please find the final version in IEEE from the link: http://ieeexplore.ieee.org/document/xxxxxxx/. Digital Object Identifier: 10.1109/xxxxxxxxx}
.%
{} %
\end{abstract}

\section{Introduction\label{sec:Introduction}}

Network densification is envisioned to be the key solution to meet
users' traffic demands in the 5th-generation (5G) networks~\cite{Tutor_smallcell}%
.%
{} Indeed, the orthogonal deployment\footnote{The orthogonal deployment means that small cells and macrocells operate
on different frequency spectrum, i.e., Small Cell Scenario \#2a~\cite{TR36.872}).} of dense small cell networks (SCNs) within the existing macrocell
networks~\cite{TR36.872} has been the workhorse for capacity enhancement
in the 4th-generation (4G), developed by the 3rd Generation Partnership
Project (3GPP), and this approach to capacity enhancement will continue
in 5G with the adoption of ultra-dense networks (UDNs)~\cite{Tutor_smallcell,Ge2016UDNs}.%
{} In this paper, we focus on the analysis of such orthogonal deployment
of UDNs.

Despite of its benefits, the SCN densification also opens up new research
questions. In particular, scheduling has been conceived as an effective
technique used at base stations (BSs) to efficiently use the available
spectrum and improve the overall system throughput. In most cases,
a proportional fair (PF) scheduler~\cite{Choi2007PF} serves as an
appealing technique that offers a good trade-off between maximizing
overall throughput and improving fairness among user equipments (UEs)
with diverse channel conditions. However, the gains of the PF scheduler
may be limited in UDNs, mostly because the number of UEs per active
BS is considerably reduced%
. This gives rise to the question of \emph{whether the PF scheduling
is as efficient for UDNs as it is for sparse networks}, or whether
it can be substituted by other ones of lower complexity, such as a
round-robin (RR) one that semi-randomly selects UEs to serve. In this
paper, we answer this fundamental question by theoretical analyses.%

To the best of our knowledge, there has been no prior work on the
theoretical study of the PF scheduler in the context of UDNs, the
density of which could be as large as tens of thousands of BSs per
$\textrm{km}^{2}$~\cite{Tutor_smallcell}. Generally speaking, the
existing work on PF schedulers does not scale well with the network
densification. In more detail, the studies in the literature can be
classified into 3 categories: \emph{(i)} analysis of a scenario with
merely one BS%
~\cite{Choi2007PF,Liu2010PFsingleBS}, \emph{(ii)} analysis of a
scenario with a limited number of BSs~\cite{Wu2011PFmultiBS,Liu2015PFmultipleBS},
which quickly becomes computationally infeasible for UDNs%
, and \emph{(iii)} system-level simulations for large-scale networks~\cite{Jafari2015scheduling},
which lacks analytical rigor.%

Compared with the existing work, the main contributions of this paper
are:
\begin{itemize}
\item For the first time, we use stochastic geometry~\cite{book_Haenggi}
to derive the analytical results of the coverage probability and the
area spectral efficiency (ASE) performance for UDNs with the PF schedulers
used at BSs. The key point of our analysis is that the typical user
is no longer a random user as assumed in most studies of stochastic
geometry~\cite{book_Haenggi}.%
\item By comparing the previous results of the RR scheduler and our new
results of the PF scheduler, we quantify the loss of the multi-user
diversity of the PF scheduler with the network densification, which
leads to the conclusion that the RR scheduler should be used in UDNs
to simplify the radio resource management (RRM).
\end{itemize}

\section{Network Scenario and System Model\label{sec:System-Model}}

In this section, we present the network scenario, the wireless system
model and the PF scheduler considered in this paper.

\subsection{Network Scenario\label{subsec:Network-Scenario}}

For a certain time-frequency resource block, we consider a downlink
(DL) cellular network with BSs deployed on a plane according to a
homogeneous Poisson point process (HPPP) $\Phi$ with a density of
$\lambda$ $\textrm{BSs/km}^{2}$. Active UEs are also Poisson distributed
in the considered DL network with a density of $\rho$ $\textrm{UEs/km}^{2}$.
Here, we only consider active UEs in the network because non-active
UEs do not trigger data transmission%
.%
{} As shown in~\cite{Tutor_smallcell}, a typical density of the active
UEs in 5G is around $300\thinspace\textrm{UEs/km}^{2}$.

In practice, a BS will enter an idle mode if there is no UE connected
to it, which reduces the interference to neighboring UEs as well as
the energy consumption of the network. The set of active BSs should
be determined by a user association strategy (UAS). In this paper,
we assume a practical UAS as in~\cite{our_work_TWC2016}, where each
UE is connected to the BS having the maximum average received signal
strength, which will be formally presented in Subsection~\ref{subsec:Wireless-System-Model}.%
{}

Note that such BS idle mode operation is not trivial, which even changes
the capacity scaling law~\cite{Ding2017capScaling}. Since UEs are
randomly and uniformly distributed in the network, we assume that
the active BSs also follow an HPPP distribution $\tilde{\Phi}$~\cite{dynOnOff_Huang2012},
the density of which is denoted by $\tilde{\lambda}$ $\textrm{BSs/km}^{2}$.
Note that $\tilde{\lambda}\leq\lambda$ and $\tilde{\lambda}\leq\rho$,
since one UE is served by at most one BS.%
{}

From~\cite{dynOnOff_Huang2012,Ding2016IMC_GC}, $\tilde{\lambda}$
is given by%
\begin{equation}
\tilde{\lambda}=\lambda\left[1-\frac{1}{\left(1+\frac{\rho}{q\lambda}\right)^{q}}\right],\label{eq:lambda_tilde_Huang}
\end{equation}
where according to~\cite{Ding2016IMC_GC}, $q$ depends on the path
loss model, which will be presented in Subsection~\ref{subsec:Wireless-System-Model}.

According to~\cite{dynOnOff_Huang2012}, %
the per-BS coverage area size $X$ can be approximately characterized
by a Gamma distribution and the probability density function (PDF)
of $X$ can be expressed as
\begin{equation}
f_{X}(x)=\left(q\lambda\right)^{q}x^{q-1}\frac{\exp(\text{\textminus}q\lambda x)}{\text{\ensuremath{\Gamma}}(q)},\label{eq:cov_area_size_PDF_Gamma}
\end{equation}
where $\text{\ensuremath{\Gamma}}(\cdot)$ is the Gamma function~\cite{Book_Integrals}.
The UE number per BS is denoted by a random variable (RV) $K$, and
the probability mass function (PMF) of $K$ can be calculated as
\begin{eqnarray}
f_{K}\left(k\right)\hspace{-0.2cm} & = & \hspace{-0.2cm}\Pr\left[K=k\right]\nonumber \\
\hspace{-0.2cm} & \overset{\left(a\right)}{=} & \hspace{-0.2cm}\int_{0}^{+\infty}\frac{\left(\rho x\right)^{k}}{k!}\exp(-\rho x)f_{X}(x)dx\nonumber \\
\hspace{-0.2cm} & \overset{\left(b\right)}{=} & \hspace{-0.2cm}\frac{\Gamma(k+q)}{\Gamma(k+1)\Gamma(q)}\left(\frac{\rho}{\rho+q\lambda}\right)^{k}\left(\frac{q\lambda}{\rho+q\lambda}\right)^{q},\label{eq:NB_PMF}
\end{eqnarray}
where $\left(a\right)$ is due to the HPPP distribution of UEs and
\emph{$\left(b\right)$ }is obtained from (\ref{eq:cov_area_size_PDF_Gamma}).
It can be seen from (\ref{eq:NB_PMF}) that $K$ follows a Negative
Binomial distribution~\cite{Book_Integrals}, i.e., $K\sim\textrm{NB}\left(q,\frac{\rho}{\rho+q\lambda}\right)$.

As discussed in Subsection~\ref{subsec:Network-Scenario}, we assume
that a BS with $K=0$ is not active%
. Thus, we focus on the active BSs and denote\emph{ the UE number
per active BS} by a positive RV $\tilde{K}$. Considering (\ref{eq:NB_PMF})%
, we can conclude that $\tilde{K}$ follows a truncated Negative Binomial
distribution%
, the PMF of which is denoted by $f_{\tilde{K}}\left(\tilde{k}\right),\tilde{k}\in\left\{ 1,2,\ldots,+\infty\right\} $
and can be written as
\begin{equation}
f_{\tilde{K}}\left(\tilde{k}\right)=\Pr\left[\tilde{K}=\tilde{k}\right]=\frac{f_{K}\left(\tilde{k}\right)}{1-f_{K}\left(0\right)}.\label{eq:truncNB_PMF}
\end{equation}
\textbf{}%
Furthermore, the cumulative mass function (CMF) of $\tilde{K}$ can
be written as
\begin{equation}
F_{\tilde{K}}\left(\tilde{k}\right)=\sum_{t=1}^{\tilde{k}}f_{\tilde{K}}\left(t\right).\label{eq:truncNB_CMF}
\end{equation}

\subsection{Wireless System Model\label{subsec:Wireless-System-Model}}

Following~\cite{our_work_TWC2016}, we adopt a general path loss
model, where the path loss $\zeta\left(r\right)$ is a multi-piece
function of $r$ written as%
\begin{equation}
\zeta\left(r\right)=\begin{cases}
\zeta_{1}\left(r\right), & \textrm{when }0\leq r\leq d_{1}\\
\zeta_{2}\left(r\right), & \textrm{when }d_{1}<r\leq d_{2}\\
\vdots & \vdots\\
\zeta_{N}\left(r\right), & \textrm{when }r>d_{N-1}
\end{cases},\label{eq:prop_PL_model}
\end{equation}
where each piece $\zeta_{n}\left(r\right),n\in\left\{ 1,2,\ldots,N\right\} $
is modeled as
\begin{equation}
\zeta_{n}\hspace{-0.1cm}\left(r\right)\hspace{-0.1cm}=\hspace{-0.1cm}\begin{cases}
\hspace{-0.2cm}\begin{array}{l}
\zeta_{n}^{\textrm{L}}\hspace{-0.1cm}\left(r\right)=A_{n}^{{\rm {L}}}r^{-\alpha_{n}^{{\rm {L}}}},\\
\zeta_{n}^{\textrm{NL}}\hspace{-0.1cm}\left(r\right)=A_{n}^{{\rm {NL}}}r^{-\alpha_{n}^{{\rm {NL}}}},
\end{array} & \hspace{-0.2cm}\hspace{-0.1cm}\hspace{-0.3cm}\begin{array}{l}
\textrm{LoS Prob.:}~\textrm{Pr}_{n}^{\textrm{L}}\left(r\right)\\
\textrm{NLoS Prob.:}~1-\textrm{Pr}_{n}^{\textrm{L}}\left(r\right)
\end{array}\hspace{-0.1cm}\hspace{-0.1cm},\end{cases}\hspace{-0.1cm}\hspace{-0.1cm}\label{eq:PL_BS2UE}
\end{equation}
where%

\begin{itemize}
\item $\zeta_{n}^{\textrm{L}}\left(r\right)$ and $\zeta_{n}^{\textrm{NL}}\left(r\right),n\in\left\{ 1,2,\ldots,N\right\} $
are the $n$-th piece path loss functions for the LoS transmission
and the NLoS transmission, respectively,
\item $A_{n}^{{\rm {L}}}$ and $A_{n}^{{\rm {NL}}}$ are the path losses
at a reference distance $r=1$ for the LoS and the NLoS cases, respectively,
\item $\alpha_{n}^{{\rm {L}}}$ and $\alpha_{n}^{{\rm {NL}}}$ are the path
loss exponents for the LoS and the NLoS cases, respectively.
\end{itemize}
Moreover, $\textrm{Pr}_{n}^{\textrm{L}}\left(r\right)$ is the $n$-th
piece LoS probability function that a transmitter and a receiver separated
by a distance $r$ has a LoS path, which is assumed to be \emph{a
monotonically decreasing function} with regard to $r$%
~\cite{TR36.828,SCM_pathloss_model}. %

As a special case to show our analytical results, we consider the
two-piece path loss and the exponential LoS probability functions
defined by the 3GPP~\cite{TR36.828}. Specifically, we have $N=2$,
$\zeta_{1}^{{\rm {L}}}\left(w\right)=\zeta_{2}^{{\rm {L}}}\left(w\right)=A^{{\rm {L}}}w^{-\alpha^{{\rm {L}}}}$,
$\zeta_{1}^{{\rm {NL}}}\left(w\right)=\zeta_{2}^{{\rm {NL}}}\left(w\right)=A^{{\rm {NL}}}w^{-\alpha^{{\rm {NL}}}}$,
$\textrm{Pr}_{1}^{{\rm {L}}}\left(w\right)=1-5\exp\left(-R_{1}/w\right)$,
and $\textrm{Pr}_{2}^{{\rm {L}}}\left(w\right)=5\exp\left(-w/R_{2}\right)$,
where $R_{1}=156$\ m, $R_{2}=30$\ m, and $d_{1}=\frac{R_{1}}{\ln10}=67.75$\ m~\cite{TR36.828}.
For clarity, this case is referred to as \textbf{the 3GPP Case} hereafter.

As discussed in Subsection~\ref{subsec:Network-Scenario}, we assume
that each UE is connected to the BS having the maximum average received
signal strength, which is equivalent to the BS with the largest $\zeta\left(r\right)$.%
{} Finally, we assume that each BS/UE is equipped with an isotropic
antenna, and that the multi-path fading between a BS and a UE is modeled
as independently identical distributed (i.i.d.) Rayleigh fading~\cite{our_work_TWC2016}.
In order to make the 3GPP Case even more practical, we will further
consider distance-dependent Rician fading~\cite{SCM_pathloss_model}%
{} in our simulations to show their minor impact on our conclusions.
More specifically, we adopt the practical Rician fading defined in
the 3GPP~\cite{SCM_pathloss_model}, where the $K$ factor in dB
unit (the ratio between the power in the direct path and the power
in the other scattered paths) is modeled as $K[\textrm{dB}]=13-0.03r$,
where $r$ is the distance in meter. %

\subsection{The PF Scheduler\label{subsec:The-PF-Scheduler}}

The original operation of the PF scheduler is as follows~\cite{Choi2007PF},
\begin{itemize}
\item First, the average throughput of each UE is tracked by an exponential
moving average at the BS.
\item Second, each UE frequently feeds back its channel state information
(CSI) to its serving BS, so that such BS can calculate the ratio of
the instantaneous achievable rate to the average throughput for each
user, which is defined as a PF metric for UE selection.
\item Finally, the UE with the maximum PF metric will be selected for DL
transmission, which is formulated as
\begin{equation}
u^{*}=\underset{u\in\left\{ 1,2,\ldots,\tilde{k}\right\} }{\arg\max}\left\{ \frac{\tilde{R}_{u}}{\bar{R}_{u}}\right\} ,\label{eq:orig_PF}
\end{equation}
where $u$, $u^{*}$, $\tilde{R}_{u}$ and $\bar{R}_{u}$ denote the
UE index, the selected UE index, the instantaneous achievable rate
of UE $u$ and the average throughput of UE $u$, respectively. Note
that the distribution of $\tilde{k}$ has been discussed in (\ref{eq:truncNB_PMF}).%
{}
\end{itemize}

From a network performance analysis point of view, it is very difficult,
if not impossible, to analyze the original PF scheduler given by (\ref{eq:orig_PF}).
This is because that the objective of a performance analysis is usually
to derive the average user throughput $\bar{R}_{u}$ or aggregate
inter-cell interference, but in this case it is part of the PF metric,
i.e., $\frac{\tilde{R}_{u}}{\bar{R}_{u}}$, and it should be known
and plugged into the UE selection criterion of (\ref{eq:orig_PF})
before the performance analysis of $\bar{R}_{u}$ is carried out.
A widely adopted approach to tackle this dilemma is to use alternative
measures of CSI in a PF metric, instead of $\tilde{R}_{u}$ and $\bar{R}_{u}$~\cite{Choi2007PF,Liu2010PFsingleBS,Wu2011PFmultiBS,Liu2015PFmultipleBS}.%

Here, we follow the framework developed in~\cite{Choi2007PF}, where
the authors proposed to use the ratio of the instantaneous signal-to-noise
ratio (SNR) to the average SNR as a PF metric instead of the original
one. More specifically, the UE selection criterion of the PF scheduler
proposed in~\cite{Choi2007PF} is given by
\begin{equation}
u^{*}=\underset{u\in\left\{ 1,2,\ldots,\tilde{k}\right\} }{\arg\max}\left\{ \frac{\tilde{Z}_{u}}{\bar{Z}_{u}}\right\} ,\label{eq:mod_PF}
\end{equation}
where $\tilde{Z}_{u}$ and $\bar{Z}_{u}$ denote the instantaneous
SNR of UE $u$ and the average SNR of UE $u$, respectively. \textbf{Although
this criterion of (\ref{eq:mod_PF}) is not exactly the same as that
of (\ref{eq:orig_PF}), it captures the important characteristics
of the PF scheduler}: \emph{(i)} allowing preference to UEs with relatively
good instantaneous channels with respect to their average ones since
\emph{$\tilde{R}_{u}$ is a strictly monotonically increasing function
of $\tilde{Z}_{u}$}, \emph{and (ii)} allocating the same portion
of resource to each UE in the long term to enforce fairness, because
the chance of $\tilde{Z}_{u}\geq\bar{Z}_{u}$ is almost the same for
all UEs. Since the accuracy and the practicality of (\ref{eq:mod_PF})
have been well established in~\cite{Choi2007PF}, we will focus on
studying the PF scheduler characterized by (\ref{eq:mod_PF})%
.

\section{Main Results\label{sec:Main-Results}}

In this section, we study the coverage probability and the ASE performance
of a typical UE located at the origin $o$.

\subsection{The Coverage Probability\label{subsec:The-Coverage-Probability}}

{\small{}}
\begin{algorithm*}
\begin{thm}
\label{thm:p_cov_UAS1}Considering the path loss model in (\ref{eq:prop_PL_model})
and the PF scheduler model in (\ref{eq:mod_PF}), we can derive $p^{{\rm {cov}}}\left(\lambda,\gamma\right)$
as
\begin{equation}
p^{{\rm {cov}}}\left(\lambda,\gamma\right)=\sum_{n=1}^{N}\left(T_{n}^{{\rm {L}}}+T_{n}^{{\rm {NL}}}\right),\label{eq:Theorem_1_p_cov}
\end{equation}
where $T_{n}^{{\rm {L}}}=\int_{d_{n-1}}^{d_{n}}\mathbb{E}_{\left[\tilde{k}\right]}\left\{ {\rm {Pr}}\left[\frac{P\zeta_{n}^{{\rm {L}}}\left(r\right)y\left(\tilde{k}\right)}{I_{{\rm {agg}}}+P_{{\rm {N}}}}>\gamma\right]\right\} f_{R,n}^{{\rm {L}}}\left(r\right)dr$,
$T_{n}^{{\rm {NL}}}=\int_{d_{n-1}}^{d_{n}}\mathbb{E}_{\left[\tilde{k}\right]}\left\{ {\rm {Pr}}\left[\frac{P\zeta_{n}^{{\rm {NL}}}\left(r\right)y\left(\tilde{k}\right)}{I_{{\rm {agg}}}+P_{{\rm {N}}}}>\gamma\right]\right\} \left(r\right)dr$,
and $d_{0}$ and $d_{N}$ are defined as $0$ and $+\infty$, respectively.
Moreover, $f_{R,n}^{{\rm {L}}}\left(r\right)$ and $f_{R,n}^{{\rm {NL}}}\left(r\right)$
$\left(d_{n-1}<r\leq d_{n}\right)$, are represented by
\begin{equation}
f_{R,n}^{{\rm {L}}}\left(r\right)=\exp\left(\hspace{-0.1cm}-\hspace{-0.1cm}\int_{0}^{r_{1}}\left(1-{\rm {Pr}}^{{\rm {L}}}\left(u\right)\right)2\pi u\lambda du\right)\exp\left(\hspace{-0.1cm}-\hspace{-0.1cm}\int_{0}^{r}{\rm {Pr}}^{{\rm {L}}}\left(u\right)2\pi u\lambda du\right){\rm {Pr}}_{n}^{{\rm {L}}}\left(r\right)2\pi r\lambda,\hspace{-0.1cm}\hspace{-0.1cm}\hspace{-0.1cm}\hspace{-0.1cm}\label{eq:geom_dis_PDF_UAS1_LoS_thm}
\end{equation}
and
\begin{equation}
\hspace{-0.1cm}\hspace{-0.1cm}\hspace{-0.1cm}\hspace{-0.1cm}\hspace{-0.1cm}\hspace{-0.1cm}f_{R,n}^{{\rm {NL}}}\left(r\right)=\exp\left(\hspace{-0.1cm}-\hspace{-0.1cm}\int_{0}^{r_{2}}{\rm {Pr}}^{{\rm {L}}}\left(u\right)2\pi u\lambda du\right)\exp\left(\hspace{-0.1cm}-\hspace{-0.1cm}\int_{0}^{r}\left(1-{\rm {Pr}}^{{\rm {L}}}\left(u\right)\right)2\pi u\lambda du\right)\left(1-{\rm {Pr}}_{n}^{{\rm {L}}}\left(r\right)\right)2\pi r\lambda,\hspace{-0.1cm}\hspace{-0.1cm}\hspace{-0.1cm}\hspace{-0.1cm}\label{eq:geom_dis_PDF_UAS1_NLoS_thm}
\end{equation}
where $r_{1}=\underset{r_{1}}{\arg}\left\{ \zeta^{{\rm {NL}}}\left(r_{1}\right)=\zeta_{n}^{{\rm {L}}}\left(r\right)\right\} $
and $r_{2}=\underset{r_{2}}{\arg}\left\{ \zeta^{{\rm {L}}}\left(r_{2}\right)=\zeta_{n}^{{\rm {NL}}}\left(r\right)\right\} $.%
\end{thm}
\begin{IEEEproof}
See Appendix~A.
\end{IEEEproof}
\end{algorithm*}
{\small{}}
\begin{algorithm*}
\begin{thm}
\label{thm:distance-specific-P-cov}Considering the truncated Negative
Binomial distribution of the UE number per active BS, $\tilde{K}$,
characterized in (\ref{eq:truncNB_PMF}), we can derive $\mathbb{E}_{\left[\tilde{k}\right]}\left\{ {\rm {Pr}}\left[\frac{P\zeta_{n}^{{\rm {L}}}\left(r\right)y\left(\tilde{k}\right)}{I_{{\rm {agg}}}+P_{{\rm {N}}}}>\gamma\right]\right\} $,
which will be used in Theorem~\ref{thm:p_cov_UAS1} as
\begin{equation}
\mathbb{E}_{\left[\tilde{k}\right]}\left\{ {\rm {Pr}}\left[\frac{P\zeta_{n}^{{\rm {L}}}\left(r\right)y\left(\tilde{k}\right)}{I_{{\rm {agg}}}+P_{{\rm {N}}}}>\gamma\right]\right\} =\sum_{\tilde{k}=1}^{\tilde{K}^{{\rm {max}}}}\left[1-\sum_{t=0}^{\tilde{k}}\left(\begin{array}{c}
\tilde{k}\\
t
\end{array}\right)\left(-\delta_{n}^{{\rm {L}}}\left(r\right)\right)^{t}\mathscr{L}_{I_{{\rm {agg}}}}^{{\rm {L}}}\left(\frac{t\gamma}{P\zeta_{n}^{{\rm {L}}}\left(r\right)}\right)\right]f_{\tilde{K}}\left(\tilde{k}\right),\label{eq:condPr_SINR_UAS1_LoS_thm}
\end{equation}
where $\tilde{K}^{{\rm {max}}}$ is a large enough integer that makes
$F_{\tilde{K}}\left(\tilde{K}^{{\rm {max}}}\right)$ in (\ref{eq:truncNB_CMF})
close to one with a gap of a small value $\epsilon$ so that the expectation
value in (\ref{eq:condPr_SINR_UAS1_LoS_thm}) can be accurately evaluated
over $\tilde{k}$, $f_{\tilde{K}}\left(\tilde{k}\right)$ is obtained
from (\ref{eq:truncNB_PMF}), $\delta_{n}^{{\rm {L}}}\left(r\right)$
is expressed by
\begin{equation}
\delta_{n}^{{\rm {L}}}\left(r\right)=\exp\left(-\frac{\gamma P_{{\rm {N}}}}{P\zeta_{n}^{{\rm {L}}}\left(r\right)}\right),\label{eq:delta_term_LoS_UAS1_general_seg_thm}
\end{equation}
and $\mathscr{L}_{I_{{\rm {agg}}}}^{{\rm {L}}}\left(s\right)$ is
the Laplace transform of $I_{{\rm {agg}}}$ for LoS signal transmission
evaluated at $s$, which can be further written as
\begin{equation}
\mathscr{L}_{I_{{\rm {agg}}}}^{{\rm {L}}}\left(s\right)=\exp\left(-2\pi\tilde{\lambda}\int_{r}^{+\infty}\frac{{\rm {Pr}}^{{\rm {L}}}\left(u\right)u}{1+\left(sP\zeta^{{\rm {L}}}\left(u\right)\right)^{-1}}du\right)\exp\left(-2\pi\tilde{\lambda}\int_{r_{1}}^{+\infty}\frac{\left[1-{\rm {Pr}}^{{\rm {L}}}\left(u\right)\right]u}{1+\left(sP\zeta^{{\rm {NL}}}\left(u\right)\right)^{-1}}du\right).\label{eq:laplace_term_LoS_UAS1_general_seg_thm}
\end{equation}

$\quad$In a similar way, $\mathbb{E}_{\left[\tilde{k}\right]}\left\{ {\rm {Pr}}\left[\frac{P\zeta_{n}^{{\rm {NL}}}\left(r\right)y\left(\tilde{k}\right)}{I_{{\rm {agg}}}+P_{{\rm {N}}}}>\gamma\right]\right\} $
is computed by
\begin{equation}
\mathbb{E}_{\left[\tilde{k}\right]}\left\{ {\rm {Pr}}\left[\frac{P\zeta_{n}^{{\rm {NL}}}\left(r\right)y\left(\tilde{k}\right)}{I_{{\rm {agg}}}+P_{{\rm {N}}}}>\gamma\right]\right\} =\sum_{\tilde{k}=1}^{\tilde{K}^{{\rm {max}}}}\left[1-\sum_{t=0}^{\tilde{k}}\left(\begin{array}{c}
\tilde{k}\\
t
\end{array}\right)\left(-\delta_{n}^{{\rm {NL}}}\left(r\right)\right)^{t}\mathscr{L}_{I_{{\rm {agg}}}}^{{\rm {NL}}}\left(\frac{t\gamma}{P\zeta_{n}^{{\rm {NL}}}\left(r\right)}\right)\right]f_{\tilde{K}}\left(\tilde{k}\right),\label{eq:condPr_SINR_UAS1_NLoS_thm}
\end{equation}
where $\delta_{n}^{{\rm {NL}}}\left(r\right)$ is expressed by
\begin{equation}
\delta_{n}^{{\rm {NL}}}\left(r\right)=\exp\left(-\frac{\gamma P_{{\rm {N}}}}{P\zeta_{n}^{{\rm {NL}}}\left(r\right)}\right),\label{eq:delta_term_NLoS_UAS1_general_seg_thm}
\end{equation}
and $\mathscr{L}_{I_{{\rm {agg}}}}^{{\rm {NL}}}\left(s\right)$ is
the Laplace transform of $I_{{\rm {agg}}}$ for NLoS signal transmission
evaluated at $s$, which can be further written as
\begin{equation}
\mathscr{L}_{I_{{\rm {agg}}}}^{{\rm {NL}}}\left(s\right)=\exp\left(-2\pi\tilde{\lambda}\int_{r_{2}}^{+\infty}\frac{{\rm {Pr}}^{{\rm {L}}}\left(u\right)u}{1+\left(sP\zeta^{{\rm {L}}}\left(u\right)\right)^{-1}}du\right)\exp\left(-2\pi\tilde{\lambda}\int_{r}^{+\infty}\frac{\left[1-{\rm {Pr}}^{{\rm {L}}}\left(u\right)\right]u}{1+\left(sP\zeta^{{\rm {NL}}}\left(u\right)\right)^{-1}}du\right).\label{eq:laplace_term_NLoS_UAS1_general_seg_thm}
\end{equation}
\end{thm}
\begin{IEEEproof}
See Appendix~B.
\end{IEEEproof}
\end{algorithm*}
{\small \par}

First, we investigate the coverage probability that the typical UE's
signal-to-interference-plus-noise ratio (SINR) is above a designated
threshold $\gamma$:
\begin{equation}
p^{\textrm{cov}}\left(\lambda,\gamma\right)=\textrm{Pr}\left[\mathrm{SINR}>\gamma\right],\label{eq:Coverage_Prob_def}
\end{equation}
where the typical UE's SINR is computed by
\begin{equation}
\mathrm{SINR}=\frac{P\zeta\left(r\right)y\left(\tilde{k}\right)}{I_{\textrm{agg}}+P_{{\rm {N}}}},\label{eq:SINR}
\end{equation}
where $y\left(\tilde{k}\right)$ is the channel gain on condition
of the UE number $\tilde{k}$%
, $P$ and $P_{{\rm {N}}}$ are the BS transmission power and the
additive white Gaussian noise (AWGN) power at each UE, respectively,
and $I_{\textrm{agg}}$ is the cumulative interference given by
\begin{equation}
I_{\textrm{agg}}=\sum_{i:\,b_{i}\in\tilde{\Phi}\setminus b_{o}}P\beta_{i}g_{i},\label{eq:cumulative_interference}
\end{equation}
where $b_{o}$ is the BS serving the typical UE, and $b_{i}$, $\beta_{i}$
and $g_{i}$ are the $i$-th interfering BS, the path loss from $b_{i}$
to the typical UE and the multi-path fading channel gain associated
with $b_{i}$ (exponentially distributed), respectively. Note that
in (\ref{eq:cumulative_interference}), only the BSs in $\tilde{\Phi}\setminus b_{o}$
inject effective interference into the network, where $\tilde{\Phi}$
denotes the set of the active BSs. %

It is very important to note that the distribution of $y\left(\tilde{k}\right)$
should be derived according to (\ref{eq:mod_PF}). More specifically,
based on the variable definition in (\ref{eq:SINR}), we can reformulate
(\ref{eq:mod_PF}) as
\begin{equation}
u^{*}=\underset{u\in\left\{ 1,2,\ldots,\tilde{k}\right\} }{\arg\max}\left\{ \frac{\frac{P\zeta\left(r\right)h_{u}}{P_{{\rm {N}}}}}{\frac{P\zeta\left(r\right)\times1}{P_{{\rm {N}}}}}\right\} =\underset{u\in\left\{ 1,2,\ldots,\tilde{k}\right\} }{\arg\max}\left\{ h_{u}\right\} ,\label{eq:mod_PF_reformulation}
\end{equation}
where $h_{u}$ is an i.i.d. RV with a \emph{unit-mean} exponential
distribution due to our consideration of Rayleigh fading mentioned
in Subsection~\ref{subsec:Wireless-System-Model}. Thus, $y\left(\tilde{k}\right)$
can be modeled as the maximum RV of $\tilde{k}$ i.i.d. exponential
RVs. The complementary cumulative distribution function (CCDF) of
$y\left(\tilde{k}\right)$ is~\cite{book_orderStatistic}
\begin{equation}
\bar{F}_{Y\left(\tilde{k}\right)}\left(y\right)={\rm {Pr}}\left[Y\left(\tilde{k}\right)>y\right]=1-\left(1-\exp\left(-y\right)\right)^{\tilde{k}}.\label{eq:CCDF_Yk_tilde}
\end{equation}
It is easy to see that ${\rm {Pr}}\left[Y\left(\tilde{k}\right)>y\right]$
increases as $\tilde{k}$ grows, which in turn improves the typical
UE's channel gain. Note that for the RR scheduler, the typical UE
is randomly selected in the BS. Consequently, we have that $\tilde{k}=1$
in (\ref{eq:CCDF_Yk_tilde})%
{} and the analytical results for RR have been derived in~\cite{Ding2016IMC_GC}.

\begin{singlespace}
Based on the path loss model in (\ref{eq:prop_PL_model}) and the
PF scheduler model in (\ref{eq:mod_PF}), we present our result of
$p^{\textrm{cov}}\left(\lambda,\gamma\right)$ in Theorem~\ref{thm:p_cov_UAS1}.%

\end{singlespace}

Digging into Theorem~\ref{thm:p_cov_UAS1} and considering the truncated
Negative Binomial distribution of the UE number per active BS, $\tilde{K}$
, we present our results on $\mathbb{E}_{\left[\tilde{k}\right]}\left\{ {\rm {Pr}}\left[\frac{P\zeta_{n}^{{\rm {L}}}\left(r\right)y\left(\tilde{k}\right)}{I_{{\rm {agg}}}+P_{{\rm {N}}}}>\gamma\right]\right\} $
and $\mathbb{E}_{\left[\tilde{k}\right]}\left\{ {\rm {Pr}}\left[\frac{P\zeta_{n}^{{\rm {NL}}}\left(r\right)y\left(\tilde{k}\right)}{I_{{\rm {agg}}}+P_{{\rm {N}}}}>\gamma\right]\right\} $
in Theorem~\ref{thm:distance-specific-P-cov}.

Plugging Theorem~\ref{thm:distance-specific-P-cov} into Theorem~\ref{thm:p_cov_UAS1},
yields our theoretical results on $p^{\textrm{cov}}\left(\lambda,\gamma\right)$.
From Theorems~\ref{thm:p_cov_UAS1} and~\ref{thm:distance-specific-P-cov},
we can draw an important and intuitive conclusion in Lemma~\ref{lem:diminishing-UE-diversity}.
\begin{lem}
\noindent \textbf{\label{lem:diminishing-UE-diversity}}The $p^{\textrm{cov}}\left(\lambda,\gamma\right)$
of the PF scheduler converges to that of the RR scheduler as \textup{$\lambda\rightarrow+\infty$}.
\vspace{-0.4cm}
\end{lem}
\begin{IEEEproof}
\begin{doublespace}
See Appendix~C.\vspace{-0.7cm}
\end{doublespace}
\end{IEEEproof}
{\small{}}
\begin{algorithm*}
\begin{thm}
\label{thm:distance-specific-P-cov-UB}$\mathbb{E}_{\left[\tilde{k}\right]}\left\{ {\rm {Pr}}\left[\frac{P\zeta_{n}^{{\rm {L}}}\left(r\right)y\left(\tilde{k}\right)}{I_{{\rm {agg}}}+P_{{\rm {N}}}}>\gamma\right]\right\} $
and $\mathbb{E}_{\left[\tilde{k}\right]}\left\{ {\rm {Pr}}\left[\frac{P\zeta_{n}^{{\rm {NL}}}\left(r\right)y\left(\tilde{k}\right)}{I_{{\rm {agg}}}+P_{{\rm {N}}}}>\gamma\right]\right\} $
can be respectively upper bounded by
\begin{equation}
\mathbb{E}_{\left[\tilde{k}\right]}\left\{ {\rm {Pr}}\left[\frac{P\zeta_{n}^{{\rm {L}}}\left(r\right)y\left(\tilde{k}\right)}{I_{{\rm {agg}}}+P_{{\rm {N}}}}>\gamma\right]\right\} \leq\sum_{\tilde{k}=1}^{\tilde{K}^{{\rm {max}}}}\left\{ 1-\left[1-\delta_{n}^{{\rm {L}}}\left(r\right)\mathscr{L}_{I_{{\rm {agg}}}}^{{\rm {L}}}\left(\frac{\gamma}{P\zeta_{n}^{{\rm {L}}}\left(r\right)}\right)\right]^{\tilde{k}}\right\} f_{\tilde{K}}\left(\tilde{k}\right),\label{eq:UB_condPr_SINR_UAS1_LoS_thm}
\end{equation}
and
\begin{equation}
\mathbb{E}_{\left[\tilde{k}\right]}\left\{ {\rm {Pr}}\left[\frac{P\zeta_{n}^{{\rm {NL}}}\left(r\right)y\left(\tilde{k}\right)}{I_{{\rm {agg}}}+P_{{\rm {N}}}}>\gamma\right]\right\} \leq\sum_{\tilde{k}=1}^{\tilde{K}^{{\rm {max}}}}\left\{ 1-\left[1-\delta_{n}^{{\rm {NL}}}\left(r\right)\mathscr{L}_{I_{{\rm {agg}}}}^{{\rm {NL}}}\left(\frac{\gamma}{P\zeta_{n}^{{\rm {NL}}}\left(r\right)}\right)\right]^{\tilde{k}}\right\} f_{\tilde{K}}\left(\tilde{k}\right).\label{eq:UB_condPr_SINR_UAS1_NLoS_thm}
\end{equation}
\end{thm}
\begin{IEEEproof}
See Appendix~D.
\end{IEEEproof}
\end{algorithm*}
{\small \par}

Although we have obtained the closed-form expressions of $p^{\textrm{cov}}\left(\lambda,\gamma\right)$
for the PF scheduler in Theorems~\ref{thm:p_cov_UAS1} and~\ref{thm:distance-specific-P-cov},
it is important to note that Theorem~\ref{thm:distance-specific-P-cov}
is computationally intensive for the case of sparse networks, where
the maximum UE number per active BS $\tilde{K}^{{\rm {max}}}$ could
be very large, leading to complex computations for $\mathscr{L}_{I_{{\rm {agg}}}}^{{\rm {L}}}\left(\frac{t\gamma}{P\zeta_{n}^{{\rm {L}}}\left(r\right)}\right)$
and $\mathscr{L}_{I_{{\rm {agg}}}}^{{\rm {NL}}}\left(\frac{t\gamma}{P\zeta_{n}^{{\rm {NL}}}\left(r\right)}\right),t\in\left\{ 0,1,\ldots,\tilde{K}^{{\rm {max}}}\right\} $
in (\ref{eq:laplace_term_LoS_UAS1_general_seg_thm}) and (\ref{eq:laplace_term_NLoS_UAS1_general_seg_thm}),
respectively. For example, when the UE density is $\rho=300\thinspace\textrm{UEs/km}^{2}$
and the BS density is $\lambda=10\thinspace\textrm{BSs/km}^{2}$,
$\tilde{K}^{{\rm {max}}}$ should be as large as 102 to make $F_{\tilde{K}}\left(\tilde{K}^{{\rm {max}}}\right)$
sufficiently close to one with a gap smaller than $\epsilon=0.001$%
. As a result, we need to calculate the integrals in (\ref{eq:laplace_term_LoS_UAS1_general_seg_thm})
and (\ref{eq:laplace_term_NLoS_UAS1_general_seg_thm}) at least 102
times for every possible value of $r$.%
{} The good news is that Theorem~\ref{thm:distance-specific-P-cov}
is very efficient for 5G UDNs, where $\tilde{K}^{{\rm {max}}}$ is
expected to be less than 10~\cite{Tutor_smallcell}. In the next
subsection, we derive alternative and more efficient expressions for
sparse networks.%

\subsection{A Low-Complexity Upper-Bound of $p^{\textrm{cov}}\left(\lambda,\gamma\right)$\label{subsec:The-Coverage-Probability-UB}}

We present upper bounds of $\mathbb{E}_{\left[\tilde{k}\right]}\left\{ {\rm {Pr}}\left[\frac{P\zeta_{n}^{{\rm {L}}}\left(r\right)y\left(\tilde{k}\right)}{I_{{\rm {agg}}}+P_{{\rm {N}}}}>\gamma\right]\right\} $
and $\mathbb{E}_{\left[\tilde{k}\right]}\left\{ {\rm {Pr}}\left[\frac{P\zeta_{n}^{{\rm {NL}}}\left(r\right)y\left(\tilde{k}\right)}{I_{{\rm {agg}}}+P_{{\rm {N}}}}>\gamma\right]\right\} $
in Theorem~\ref{thm:distance-specific-P-cov-UB}.

The proposed upper bounds in Theorem~\ref{thm:distance-specific-P-cov-UB}
require to calculate the integrals in (\ref{eq:laplace_term_LoS_UAS1_general_seg_thm})
and (\ref{eq:laplace_term_NLoS_UAS1_general_seg_thm}) \emph{only
once} for every possible value of $r$, which makes the analysis of
sparse networks very efficient.%
{} Consequently, plugging Theorem~\ref{thm:distance-specific-P-cov-UB}
into Theorem~\ref{thm:p_cov_UAS1}, yields our theoretical results
on an upper bound of $p^{\textrm{cov}}\left(\lambda,\gamma\right)$,
which is particularly useful for sparse networks.

\subsection{The Area Spectral Efficiency\label{subsec:The-Area-Spectral}}

We also investigate the area spectral efficiency (ASE) performance
in $\textrm{bps/Hz/km}^{2}$, which is defined as~\cite{our_work_TWC2016}
\begin{equation}
A^{{\rm {ASE}}}\left(\lambda,\gamma_{0}\right)=\tilde{\lambda}\int_{\gamma_{0}}^{+\infty}\log_{2}\left(1+\gamma\right)f_{\mathit{\Gamma}}\left(\lambda,\gamma\right)d\gamma,\label{eq:ASE_def}
\end{equation}
where $\gamma_{0}$ is the minimum working SINR in a practical SCN,
and $f_{\mathit{\Gamma}}\left(\lambda,\gamma\right)$ is the PDF of
the SINR $\gamma$ observed at the typical UE for a particular value
of $\lambda$. Based on the definition of $p^{{\rm {cov}}}\left(\lambda,\gamma\right)$
in (\ref{eq:Coverage_Prob_def}) and the partial integration theorem~\cite{Book_Integrals},
(\ref{eq:ASE_def}) can be reformulated as%
\begin{eqnarray}
A^{{\rm {ASE}}}\left(\lambda,\gamma_{0}\right)\hspace{-0.1cm}\hspace{-0.1cm} & = & \hspace{-0.1cm}\hspace{-0.1cm}\frac{\tilde{\lambda}}{\ln2}\int_{\gamma_{0}}^{+\infty}\frac{p^{{\rm {cov}}}\left(\lambda,\gamma\right)}{1+\gamma}d\gamma\nonumber \\
\hspace{-0.1cm}\hspace{-0.1cm} &  & \hspace{-0.1cm}\hspace{-0.1cm}+\tilde{\lambda}\log_{2}\left(1+\gamma_{0}\right)p^{{\rm {cov}}}\left(\lambda,\gamma_{0}\right).\label{eq:ASE_def_reform}
\end{eqnarray}

\section{Simulation and Discussion\label{sec:Simulation-and-Discussion}}

In this section, we investigate network performance and use numerical
results to validate the accuracy of our analysis. According to Tables
A.1-3, A.1-4 and A.1-7 of~\cite{TR36.828}%
, we adopt the following parameters for the 3GPP Case: $\alpha^{{\rm {L}}}=2.09$,
$\alpha^{{\rm {NL}}}=3.75$, $A^{{\rm {L}}}=10^{-10.38}$, $A^{{\rm {NL}}}=10^{-14.54}$%
, %
$P=24$\ dBm, $P_{{\rm {N}}}=-95$\ dBm (including a noise figure
of 9\ dB at each UE). Besides, the UE density $\rho$ is set to $300\thinspace\textrm{UEs/km}^{2}$,
which leads to $q=4.05$ in (\ref{eq:lambda_tilde_Huang}) and (\ref{eq:cov_area_size_PDF_Gamma})~\cite{Ding2016IMC_GC}.%

\subsection{The Coverage Probability Performance of the 3GPP Case\label{subsec:Sim-p-cov-3GPP-Case}}

In Fig.~\ref{fig:p_cov_vs_lambda_gamma0dB_UEdensity300_PF}, we plot
the results of $p^{\textrm{cov}}\left(\lambda,\gamma\right)$ with
the PF scheduler for the 3GPP Case when $\rho=300\,\textrm{UEs/km}^{2}$
and $\gamma=0\,\textrm{dB}$. Note that our analytical results on
the exact performance are obtained from Theorems~\ref{thm:p_cov_UAS1}
and~\ref{thm:distance-specific-P-cov}. In contrast, our analytical
results on the upper-bound performance are obtained from Theorems~\ref{thm:p_cov_UAS1}
and~\ref{thm:distance-specific-P-cov-UB}. As a benchmark, we also
provide simulation results, and display the analytical results of
the RR scheduler reported in~\cite{Ding2016IMC_GC}. Moreover, we
show the ratio of the simulated $p^{\textrm{cov}}\left(\lambda,\gamma\right)$
of the PF scheduler to that of the RR scheduler in Fig.~\ref{fig:p_cov_vs_lambda_gamma0dB_UEdensity300_PF_over_RR}.

From these two figures, we can observe that:
\begin{itemize}
\item As can be seen from Fig.~\ref{fig:p_cov_vs_lambda_gamma0dB_UEdensity300_PF},
our analytical results well match the simulation results, which validates
the accuracy of Theorems~\ref{thm:p_cov_UAS1} and~\ref{thm:distance-specific-P-cov}.
However, as discussed in Subsection~\ref{subsec:The-Coverage-Probability},
the evaluation of our analytical results is only efficient for dense
and ultra-dense networks. Thus, in Fig.~\ref{fig:p_cov_vs_lambda_gamma0dB_UEdensity300_PF},
we are only able to show the results of $p^{\textrm{cov}}\left(\lambda,\gamma\right)$
for $\lambda\geq100\,\textrm{BSs/km}^{2}$. When $\lambda<100\,\textrm{BSs/km}^{2}$,
the proposed upper-bound results in Theorems~\ref{thm:p_cov_UAS1}
and~\ref{thm:distance-specific-P-cov-UB} successfully capture the
qualitative performance trend of the PF scheduler, with a maximum
error of 0.04 in terms of $p^{\textrm{cov}}\left(\lambda,\gamma\right)$
for sparse networks.
\begin{figure}
\noindent \begin{centering}
\includegraphics[width=8cm]{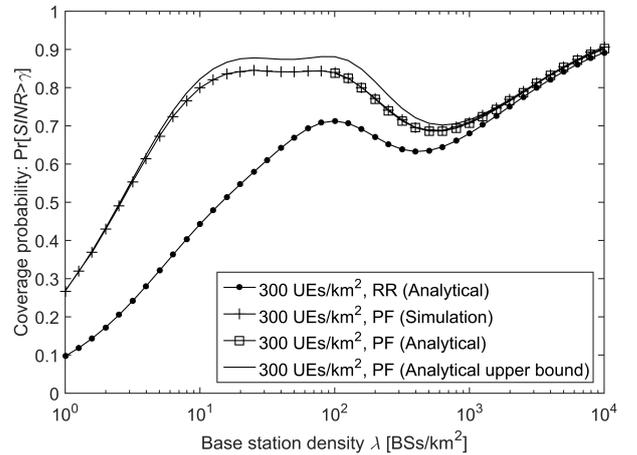}\renewcommand{\figurename}{Fig.}\caption{\label{fig:p_cov_vs_lambda_gamma0dB_UEdensity300_PF}The coverage
probability $p^{\textrm{cov}}\left(\lambda,\gamma\right)$ vs. $\lambda$
for the 3GPP Case with Rayleigh fading ($\rho=300\,\textrm{UEs/km}^{2}$,
$\gamma=0\,\textrm{dB}$). }
\par\end{centering}
\vspace{-0.4cm}
\end{figure}
\begin{figure}
\noindent \begin{centering}
\includegraphics[width=8cm]{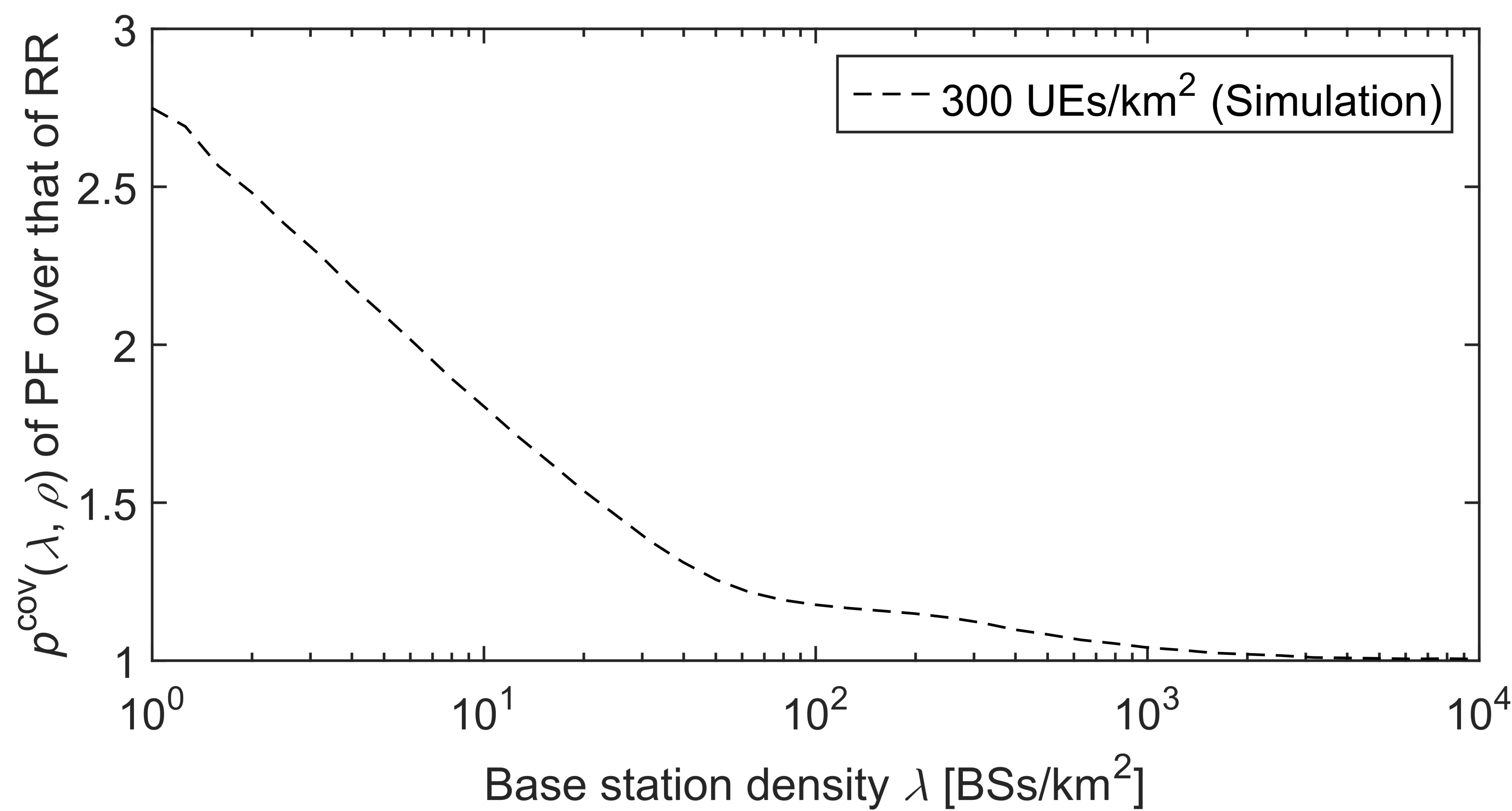}\renewcommand{\figurename}{Fig.}\caption{\label{fig:p_cov_vs_lambda_gamma0dB_UEdensity300_PF_over_RR}The ratio
of the simulated $p^{\textrm{cov}}\left(\lambda,\gamma\right)$ of
PF to that of RR for the 3GPP Case with Rayleigh fading ($\rho=300\,\textrm{UEs/km}^{2}$,
$\gamma=0\,\textrm{dB}$). }
\par\end{centering}
\vspace{-0.5cm}
\end{figure}
\item As predicted in Lemma~\ref{lem:diminishing-UE-diversity}, although
the PF scheduler shows a better performance than the RR one for all
BS densities, such performance gain diminishes as the network evolves
into an UDN due to the loss of multi-user diversity. As can be seen
from Fig.~\ref{fig:p_cov_vs_lambda_gamma0dB_UEdensity300_PF_over_RR},
the performance gain of the PF scheduler continuously decreases from
around 175$\,$\% (ratio=2.75) when $\lambda=1\,\textrm{BSs/km}^{2}$
toward zero (ratio=1) in UDNs, e.g., $\lambda=10^{4}\,\textrm{BSs/km}^{2}$.
\item The detailed explanation of the performance behavior in Fig.~\ref{fig:p_cov_vs_lambda_gamma0dB_UEdensity300_PF}
is provided as follows:
\begin{itemize}
\item When $\lambda\in\left[10^{0},10^{1}\right]\,\textrm{BSs/km}^{2}$,
the network is noise-limited, and thus the coverage probabilities
of both RR and PF increase with the BS density $\lambda$ as the network
is lightened up with more BSs.
\item When $\lambda\in\left[10^{1},10^{2}\right]\,\textrm{BSs/km}^{2}$,
$p^{\textrm{cov}}\left(\lambda,\gamma\right)$ of the PF scheduler
shows an interesting flat trail. This is because \emph{(i)} the signal
power is enhanced by LoS transmissions, as shown by the $p^{\textrm{cov}}\left(\lambda,\gamma\right)$
of the RR scheduler in that BS density region; while \emph{(ii)} the
multi-user diversity decreases in that BS density region as exhibited
in Fig.~\ref{fig:p_cov_vs_lambda_gamma0dB_UEdensity300_PF_over_RR};
and \emph{(iii)} the above two factors%
{} roughly cancel each other out.
\item When $\lambda\in\left[10^{2},10^{3}\right]\,\textrm{BSs/km}^{2}$,
the coverage probabilities of both RR and PF decrease with $\lambda$,
as the network is pushed into the interference-limited region, and
this performance degradation is due to the transition of a large number
of interfering paths from NLoS to LoS, which accelerates the growth
of the aggregate inter-cell interference~\cite{our_work_TWC2016}.
\item When $\lambda>10^{3}\,\textrm{BSs/km}^{2}$, the coverage probabilities
of both RR and PF continuously increase~\cite{Ding2016IMC_GC}.%
{} Such performance behavior can be attributed to the BS idle mode operations,
i.e., \emph{(i) }the signal power continues increasing with the network
densification, and \emph{(ii) }the interference power is controlled
because not all BSs are turned on and emit interference.%
\end{itemize}
\end{itemize}

\subsection{The Performance Impact of Rician Fading\label{subsec:perfm_3GPP_Case2}}

\begin{figure}
\noindent \begin{centering}
\includegraphics[width=8cm]{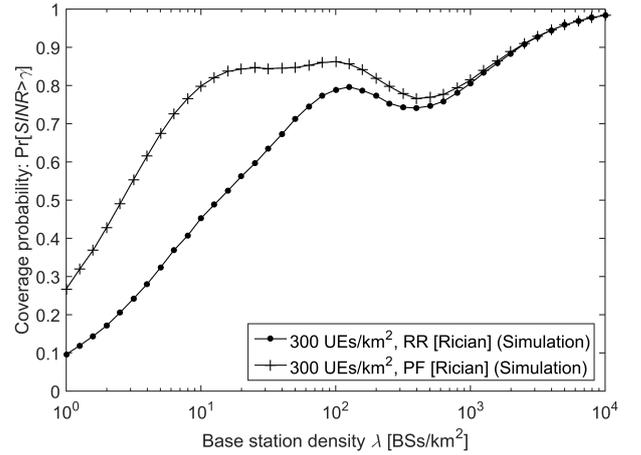}\renewcommand{\figurename}{Fig.}\caption{\label{fig:p_cov_vs_lambda_gamma0dB_UEdensity300_PF_Ric}The coverage
probability $p^{\textrm{cov}}\left(\lambda,\gamma\right)$ vs. $\lambda$
for the 3GPP Case with distance-dependent Rician fading ($\rho=300\,\textrm{UEs/km}^{2}$,
$\gamma=0\,\textrm{dB}$). }
\par\end{centering}
\vspace{-0.4cm}
\end{figure}

In this subsection, we investigate the performance for the 3GPP Case
with \textbf{distance-dependent Rician fading}%
, which has been introduced in Subsection~\ref{subsec:Wireless-System-Model}.
Due to its complex modeling, we conduct simulations to investigate
this enhanced 3GPP Case, and the results are plotted in Fig.~\ref{fig:p_cov_vs_lambda_gamma0dB_UEdensity300_PF_Ric}.
\begin{figure}
\noindent \begin{centering}
\includegraphics[width=8cm]{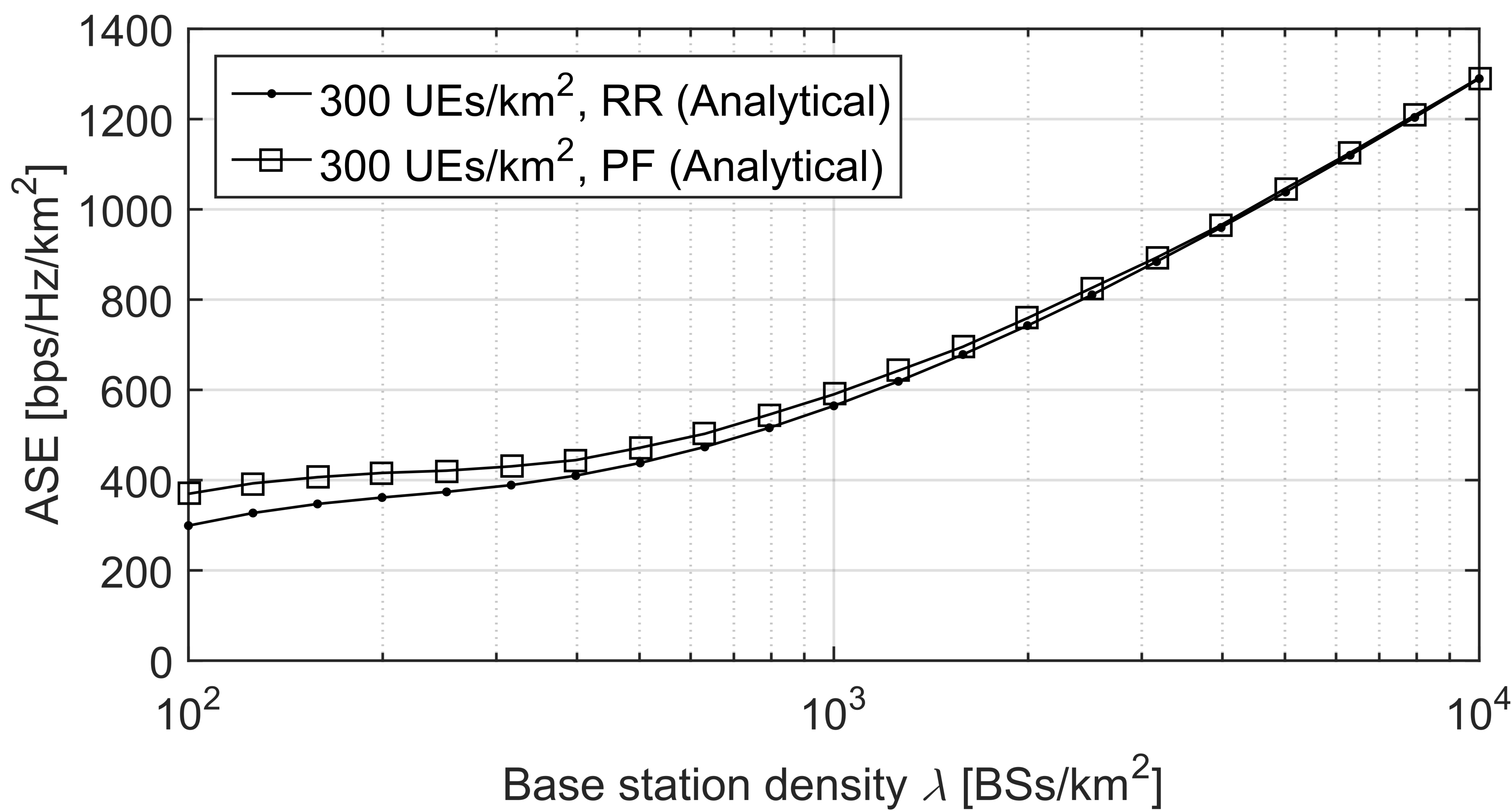}\renewcommand{\figurename}{Fig.}\caption{\label{fig:ASE_vs_lambda_gamma0dB_UEdensity300_PF}The ASE $A^{{\rm {ASE}}}\left(\lambda,\gamma_{0}\right)$
vs. $\lambda$ for the 3GPP Case with Rayleigh fading ($\rho=300\,\textrm{UEs/km}^{2}$,
$\gamma_{0}=0\,\textrm{dB}$). }
\par\end{centering}
\vspace{-0.5cm}
\end{figure}

All the conclusions in Subsections~\ref{subsec:Sim-p-cov-3GPP-Case}
are qualitatively valid for Fig.~\ref{fig:p_cov_vs_lambda_gamma0dB_UEdensity300_PF_Ric},
which shows the usefulness of our analysis and a less urgency to consider
Rician fading in this performance analysis.%
{} Note that comparing Fig.~\ref{fig:p_cov_vs_lambda_gamma0dB_UEdensity300_PF_Ric}
with Fig.~\ref{fig:p_cov_vs_lambda_gamma0dB_UEdensity300_PF}, we
can see that Rician fading closes the performance gap between PF and
RR more quickly than Rayleigh fading, due to its less variance in
channel fluctuation for the PF scheduler to exploit. %

\subsection{The ASE Performance of the 3GPP Case\label{subsec:Sim-ASE-3GPP-Case}}

Finally, we investigate the ASE performance, which is calculated from
the results of $p^{\textrm{cov}}\left(\lambda,\gamma\right)$ using
(\ref{eq:ASE_def}). Since the analytical results of $p^{\textrm{cov}}\left(\lambda,\gamma\right)$
obtained from Theorems~\ref{thm:p_cov_UAS1} and~\ref{thm:distance-specific-P-cov}
have been validated in Fig.~\ref{fig:p_cov_vs_lambda_gamma0dB_UEdensity300_PF},
we only plot the analytical results of $A^{{\rm {ASE}}}\left(\lambda,\gamma_{0}\right)$
for the PF and RR schedulers in Fig.~\ref{fig:ASE_vs_lambda_gamma0dB_UEdensity300_PF}%
. Note that we only investigate dense and ultra-dense SCNs with $\lambda\geq100\,\textrm{BSs/km}^{2}$
in Fig.~\ref{fig:ASE_vs_lambda_gamma0dB_UEdensity300_PF} because
they are the focus of 5G and Theorem~\ref{thm:distance-specific-P-cov}
is computationally efficient to evaluate them.%

As can be seen from this figure, the ASE performance of the PF scheduler
quickly converges to that of the RR scheduler in UDNs, e.g., when
$\lambda=10^{3}\,\textrm{BSs/km}^{2}$, the ASE of the PF scheduler
is around $590.1\,\textrm{bps/Hz/km}^{2}$, which is merely 4.52$\,$\%
higher than that of the RR scheduler around $564.6\,\textrm{bps/Hz/km}^{2}$.
Such gain becomes practically zero as $\lambda$ further increases.

\section{Conclusion\label{sec:Conclusion}}

In this paper, we studied the network performance of the PF scheduler.
Analytical results that are computationally efficient have been derived
for dense and ultra-dense SCNs. For sparse networks, easy-to-compute
analytical results of an upper-bound performance have been obtained
to achieve a balance between accuracy and efficiency. Considering
the negligible gain of the more complex channel-dependent PF scheduling
in UDNs,%
{} it is recommended to adopt simpler scheduling mechanisms such as
the RR scheduler to simplify the RRM, and thus reduce network complexity
for UDNs. As our future work, a non-full-buffer traffic model~\cite{Ding2016dynTDD,Zhong2017scheduling}
will be studied.%

\section*{Appendix~A: Proof of Theorem~\ref{thm:p_cov_UAS1}\label{sec:Appendix-newA}}

\begin{singlespace}

\end{singlespace}

Due to the page limit, we only provide the proof sketch of Theorem~\ref{thm:p_cov_UAS1}
as follows. In (\ref{eq:Theorem_1_p_cov}), $T_{n}^{{\rm {L}}}$ and
$T_{n}^{{\rm {NL}}}$ are the components of the coverage probability
for the case when the signal comes from \emph{the $n$-th piece LoS
path} and for the case when the signal comes from \emph{the $n$-th
piece NLoS path}, respectively. The calculation of $T_{n}^{{\rm {L}}}$
is based on (\ref{eq:geom_dis_PDF_UAS1_LoS_thm}), in which $f_{R,n}^{{\rm {L}}}\left(r\right)$
characterizes the geometrical density function of the typical UE with
\emph{no other LoS BS} and \emph{no NLoS BS} providing a better link
to the typical UE than its serving BS (a BS with \emph{the $n$-th
piece LoS path}). Besides, $\mathbb{E}_{\left[\tilde{k}\right]}\left\{ {\rm {Pr}}\left[\frac{P\zeta_{n}^{{\rm {L}}}\left(r\right)y\left(\tilde{k}\right)}{I_{{\rm {agg}}}+P_{{\rm {N}}}}>\gamma\right]\right\} $
gives the expected coverage probability over all possible values of
$\tilde{k}$ on condition of $r$. The logic of the calculation of
$T_{n}^{{\rm {NL}}}$ is similar to that of $T_{n}^{{\rm {L}}}$.

\section*{Appendix~B: Proof of Theorem~\ref{thm:distance-specific-P-cov}\label{sec:Appendix-newB}}

\begin{singlespace}

\end{singlespace}

Due to the page limit, here we only provides the key of the proof
for Theorem~\ref{thm:distance-specific-P-cov}. The derivation of
(\ref{eq:condPr_SINR_UAS1_LoS_thm}) is as follows,\vspace{0.3cm}

\noindent $\mathbb{E}_{\left[\tilde{k}\right]}\left\{ {\rm {Pr}}\left[\frac{P\zeta_{n}^{{\rm {L}}}\left(r\right)y\left(\tilde{k}\right)}{I_{{\rm {agg}}}+P_{{\rm {N}}}}>\gamma\right]\right\} $

\noindent
\begin{eqnarray}
\hspace{-0.1cm}\hspace{-0.1cm}\hspace{-0.1cm}\hspace{-0.1cm}\hspace{-0.1cm}\hspace{-0.1cm}\hspace{-0.1cm}\hspace{-0.1cm}\hspace{-0.1cm}\hspace{-0.1cm}\hspace{-0.1cm} & = & \hspace{-0.1cm}\hspace{-0.1cm}\hspace{-0.1cm}\mathbb{E}_{\left[\tilde{k}\right]}\left\{ {\rm {Pr}}\left[y\left(\tilde{k}\right)>\frac{\gamma\left(I_{{\rm {agg}}}+P_{{\rm {N}}}\right)}{P\zeta_{n}^{{\rm {L}}}\left(r\right)}\right]\right\} \nonumber \\
\hspace{-0.1cm}\hspace{-0.1cm}\hspace{-0.1cm}\hspace{-0.1cm}\hspace{-0.1cm}\hspace{-0.1cm}\hspace{-0.1cm}\hspace{-0.1cm}\hspace{-0.1cm}\hspace{-0.1cm}\hspace{-0.1cm} & \overset{(a)}{=} & \hspace{-0.1cm}\hspace{-0.1cm}\hspace{-0.1cm}\mathbb{E}_{\left[\tilde{k},I_{{\rm {agg}}}\right]}\left\{ \bar{F}_{Y\left(\tilde{k}\right)}\left(\frac{\gamma\left(I_{{\rm {agg}}}+P_{{\rm {N}}}\right)}{P\zeta_{n}^{{\rm {L}}}\left(r\right)}\right)\right\} \nonumber \\
\hspace{-0.1cm}\hspace{-0.1cm}\hspace{-0.1cm}\hspace{-0.1cm}\hspace{-0.1cm}\hspace{-0.1cm}\hspace{-0.1cm}\hspace{-0.1cm}\hspace{-0.1cm}\hspace{-0.1cm}\hspace{-0.1cm} & \overset{(b)}{=} & \hspace{-0.1cm}\hspace{-0.1cm}\hspace{-0.1cm}\sum_{\tilde{k}=1}^{\tilde{K}^{{\rm {max}}}}\hspace{-0.1cm}\mathbb{E}_{\left[I_{{\rm {agg}}}\right]}\hspace{-0.1cm}\left\{ \hspace{-0.1cm}1\hspace{-0.1cm}-\hspace{-0.1cm}\sum_{t=0}^{\tilde{k}}\hspace{-0.1cm}\left(\hspace{-0.1cm}\hspace{-0.1cm}\begin{array}{c}
\tilde{k}\\
t
\end{array}\hspace{-0.1cm}\hspace{-0.1cm}\right)\hspace{-0.1cm}\left(-\delta_{n}^{{\rm {L}}}\left(r\right)\right)^{t}\hspace{-0.1cm}\exp\left(-sI_{{\rm {agg}}}\right)\hspace{-0.1cm}\right\} \hspace{-0.1cm}f_{\tilde{K}}(\tilde{k}),\hspace{-0.1cm}\label{eq:condPr_SINR_on_r_UAS1_LoS}
\end{eqnarray}
where the step (a) of (\ref{eq:condPr_SINR_on_r_UAS1_LoS}) comes
from (\ref{eq:CCDF_Yk_tilde}), and in the step (b) of (\ref{eq:condPr_SINR_on_r_UAS1_LoS})
$s=\frac{t\gamma}{P\zeta_{n}^{{\rm {L}}}\left(r\right)}$ and $\mathbb{E}_{\left[I_{{\rm {agg}}}\right]}\left\{ \exp\left(-sI_{{\rm {agg}}}\right)\right\} =\mathscr{L}_{I_{{\rm {agg}}}}^{{\rm {L}}}\left(s\right)$
should be further plugged into (\ref{eq:condPr_SINR_on_r_UAS1_LoS})
to obtain (\ref{eq:condPr_SINR_UAS1_LoS_thm}). The calculation of
$\mathscr{L}_{I_{{\rm {agg}}}}^{{\rm {L}}}\left(s\right)$ can be
referred to~\cite{Ding2016IMC_GC}. The derivation of (\ref{eq:condPr_SINR_UAS1_NLoS_thm})
is very similar to (\ref{eq:condPr_SINR_on_r_UAS1_LoS}), which is
omitted for brevity.

\section*{Appendix~C: Proof of Lemma~\ref{lem:diminishing-UE-diversity}\label{sec:Appendix-newC}}

\begin{singlespace}
\vspace{-0.1cm}

\end{singlespace}

The key of the proof for Lemma~\ref{lem:diminishing-UE-diversity}
lies in (\ref{eq:condPr_SINR_UAS1_LoS_thm}) and (\ref{eq:condPr_SINR_UAS1_NLoS_thm})
of Theorem~\ref{thm:distance-specific-P-cov}. When $\lambda\rightarrow+\infty$,
we have that $\tilde{K}^{{\rm {max}}}\rightarrow1$ and $f_{\tilde{K}}(1)=1$%
. Hence, Theorem~\ref{thm:distance-specific-P-cov} will degenerate
to the results for the RR scheduler addressed in~\cite{Ding2016IMC_GC}.\vspace{-0.1cm}

\section*{Appendix~D: Proof of Theorem~\ref{thm:distance-specific-P-cov-UB}\label{sec:Appendix-newD}}

\begin{singlespace}
\vspace{-0.1cm}

\end{singlespace}

The key of the proof for Theorem~\ref{thm:distance-specific-P-cov-UB}
lies in using Jensen's inequality as follows~\cite{Book_Integrals}%
\begin{equation}
\mathbb{E}_{\left[I_{{\rm {agg}}}\right]}\hspace{-0.1cm}\left\{ 1\hspace{-0.1cm}-\hspace{-0.1cm}\left(1\hspace{-0.1cm}-\hspace{-0.1cm}\exp\left(-x\right)\right)^{\tilde{k}}\right\} \hspace{-0.1cm}\leq\hspace{-0.1cm}1\hspace{-0.1cm}-\hspace{-0.1cm}\left(1\hspace{-0.1cm}-\hspace{-0.1cm}\mathbb{E}_{\left[I_{{\rm {agg}}}\right]}\hspace{-0.1cm}\left\{ \exp\left(-x\right)\right\} \right)^{\tilde{k}}\hspace{-0.1cm}\hspace{-0.1cm}.\hspace{-0.1cm}\label{eq:Jensen_ineq}
\end{equation}
\vspace{-0.4cm}

\bibliographystyle{IEEEtran}
\bibliography{Ming_library}

\end{document}